\documentclass{article}
\usepackage[T1]{fontenc}
\usepackage[utf8]{inputenc}
\usepackage{ismir} % Remove the "submission" option for camera-ready version
\usepackage{amsmath,cite,url}
\usepackage{graphicx}
\usepackage{color}
\usepackage{multirow}

\newcommand{\MASK}{\texttt{MASK}}

\title{The Florence Price Art Song Dataset and Piano Accompaniment Generator}

\threeauthors
  {Tao-Tao He} {Vanderbilt University \\ \texttt{tao-tao.he@vanderbilt.edu}}
  {Martin E.\ Malandro} {Sam Houston State University \\ \texttt{malandro@shsu.edu}}
  {Douglas Shadle} {Vanderbilt University \\ \texttt{douglas.shadle@vanderbilt.edu}}

\def\authorname{T. He, M. E. Malandro, and D. Shadle}

\begin{document}

\maketitle

\begin{abstract}
Florence B. Price was a composer in the early 20\textsuperscript{th} century whose music reflects her upbringing in the American South, her African heritage, and her Western classical training. She is noted as the first African-American woman to have a symphony performed by a major orchestra. Her music has recently received renewed attention from both the public and the research community, decades after her death. In addition to other genres, Price was a prolific composer for solo voice and piano. Music historians have documented the existence of 134 art songs and piano/voice arrangements for spirituals and folk songs written by Price. We release a digital catalog of 112 of these works in MuseScore, MusicXML, MIDI, and PDF format. We also use this dataset to fine-tune a symbolic music generation model to generate accompaniments to melodies, and we conduct a blind listening experiment that shows that accompaniments generated by our model are perceived as being reflective of Florence Price's style more frequently than accompaniments generated by a baseline model. We release our model as the Florence Price Piano Accompaniment Generator alongside our dataset.\footnote{https://github.com/TT515/Florence\_Price\_Art\_Song\_Dataset}

% Through a listening test comparing accompaniments generated by a baseline model and the model fine-tuned on Price's songs, we conclude the fine-tuned model is capable of generating accompaniments in Price's style. The Florence Price Piano Accompaniment Generator can be accessed here (link redacted for review).
\end{abstract}

\section{Introduction}\label{sec:introduction}

Florence B. Price (April 9, 1888--June 3, 1953)  was a prolific % and important 
composer of classical music during the first half of the twentieth century. Born and raised in Little Rock, Arkansas, she studied piano, organ, and composition at the New England Conservatory, where she graduated with honors in 1906. For the next two decades, she primarily lived in Arkansas, where she worked principally as a music educator in the region's segregated academics. She relocated to Chicago in 1927, where she encountered a supportive community of African American musicians, artists, dancers, and writers that sustained her creative energies. Her career as a composer reached its first high point with the premiere of her award-winning \textit{Symphony in E Minor} by the Chicago Symphony Orchestra in 1933---the first time that a major American orchestra had performed music by an African American woman. During the final two decades of her life, other major performers and ensembles, such as the contralto Marian Anderson and the US Marine Band, championed her music while her reputation grew internationally through extensive publication of her shorter piano works.  \cite{brown2020heart, ege2021florence}.

Despite the attention and acclaim Price had received during her lifetime, much of her compositional output fell into obscurity after her death because many of her most important manuscripts seemingly disappeared. In 2009, however, hundreds of documents from Price's life were discovered in her old summer cottage in Kankakee County, Illinois, including letters, diaries, photographs, and manuscript scores \cite{kaufman2018lostscores}. This discovery sparked interest from researchers, who now had new information to piece together Price’s life and music, and performing musicians, who now had greater access to her extensive catalog. % {\color{red} As more of her works are now being recorded, published, and performed, researchers and the public hope to gain a better understanding of her life and her unique style.} 
Even so, access to her scores has remained limited since most of her works remained unpublished and most manuscript scores were under US copyright protection until 2024.

Price exhibited a special interest in the art song genre. % with a catalog well over one hundred unique examples. This corpus comprises some of her most stylistically distinctive works, given the wide-ranging subject matter of the texts she chose for musical settings.
Art songs are vocal compositions that set poetry to music, typically written for solo voice and piano, and composed within the Western classical tradition \cite{alamomusic}. Price's corpus of over one hundred art songs incorporates texts with diverse subject matters with her stylistically distinctive music. Price also arranged a variety of African American folk songs for solo voice and piano, % many of which were
including many spirituals--a genre of African American religious folk song originating in the US South under enslavement.

In this paper, we create and release a dataset of 112 of Price's art songs and folksong arrangements. Making this corpus available will enhance public access to Price's musical expression, musicological research on Price's life and compositional style, and technological experiments that engage with the dataset in generative ways.

% {\color{red} During the same period in which Price's music has become more prominent, research in AI-generated music has accelerated. In attempts to create generative music models that can help music creators’ creative processes (as opposed to models that perform the entire music-making process), researchers have explored models for carrying out various infilling tasks. Infilling refers to generating missing portions of music conditioned on surrounding or nearby musical context \cite{musictransformer, CA, CA2, accomontage, MMM, MMM2, midigpt, musiac,  MMMeval, pianoinpainting}. Accompaniment generation, where a melody track conditions the accompaniment tracks generated, is a type of infilling task. }

% Art songs are vocal compositions that set poetry to music, typically written for solo voice and piano, and composed within the Western classical tradition \cite{alamomusic}. The songs are a natural fit for studying relations between melody and accompaniment because of the instrumentation, and Price's catalog of such songs offers the opportunity to study how a specific composer approached these relations.

We engage in one such experiment by fine-tuning a generative symbolic music model on our dataset to create a model that generates accompaniments that are reflective of Price’s style. The model accepts melodies and outputs accompaniments to each input melody. The instrumentation of art songs make them naturally fit for studying relations between melody and accompaniment, and Price's catalog of such songs offers the opportunity to study how a specific composer approached these relations.

Our model has applications in education, performance, and research. 
% The model is fine-tuned to capture Price’s musical language, and its ability to generate accompaniments offers researchers a novel lens through which to analyze her stylistic tendencies. 
The model's ability to generate accompaniments offers researchers a novel lens for analyzing Price's stylistic tendencies. 
The model could generate piano parts for Price's incomplete works, similar to how an AI model helped complete an extended version of Schubert's "Unfinished Symphony" \cite{davis2019schubert}. It can also serve as a tool for inspiration for composers and composition students who want to incorporate elements of Price's style. For AI music researchers, it serves as a case study for model personalization---namely, how well a model can capture a composer's style based on limited data.

Our main contributions are: (1) We release a novel dataset of 112 of Price’s art songs in MuseScore, MusicXML, MIDI, and PDF format. This is the most complete digital collection of Florence Price’s vocal music works to date. 
% We include human annotations of section boundaries for all songs in the dataset. 
(2) We train and release the Florence Price Piano Accompaniment Generator, which generates piano accompaniments reflective of the style of Florence Price. (3) We conduct a blind listening experiment that shows that accompaniments generated by our model are perceived as being reflective of Florence Price's style more frequently than accompaniments generated by a baseline model.
% We demonstrate via a blind listening study that the model is perceived as writing accompaniments in Price’s style.

\section{Related Work}\label{sec:related}

Sheet music has been digitized into computer-readable formats in numerous ways, including ABC notation, music21 \cite{music21}, MIDI, and MusicXML. Digitized classical music datasets include the OpenScore Lieder Corpus \cite{OSLieder}, which includes 1356 works of vocal music (art songs) by classical music composers (primarily in the Romantic Era), and the OpenScore String Quartet Corpus \cite{OSString}. Our dataset adds to collection of composer-specific catalogs, such as the JSB Chorales dataset \cite{jsbach_chorales}, which encodes 382 of J.S. Bach’s four-part chorales in various formats and is widely used in machine learning contexts---see, e.g., \cite{deepbach, coco}.

% Florence Price's first biography, written by Rae Linda Brown, was published in 2020 \cite{brown2020heart}.
The in-progress Florence B. Price Works Catalog \cite{price_catalog}, launched in 2023, keeps track of information about all documented compositions by Price, including where each composition's existing manuscripts can be found. The most complete collection of Florence Price’s vocal music to date is {\em 44 Art Songs and Spirituals} \cite{44songs}, published in 2015. More of her recently discovered songs have been recorded and released, including Karen Slack’s Grammy\textsuperscript{\textregistered}-winning album Beyond the Years: Unpublished Songs of Florence Price \cite{slack2024beyond}, containing 19 unpublished songs by Price and released in 2024. However, until now, most of the sheet music to these songs has remained unpublished. Prior to this work, there were no digital repositories of Price's works in symbolic format---vocal or otherwise.

Transformers have been widely used for symbolic music generation since Google’s Music Transformer release in 2018 \cite{musictransformer}. The task of symbolic music infilling, and specifically accompaniment generation, has been explored in \cite{musictransformer} as well as in Composer's Assistant \cite{CA, CA2}, Accomontage \cite{accomontage}, and others \cite{MMM, MMM2, midigpt, musiac}. In ShredGP \cite{ShredGP} and ProgGP \cite{ProgGP}, it is shown that transformer models can be fine-tuned on small datasets to train the models to write in specific musical styles. However, fine-tuning for style-conditioned {\em accompaniment} generation is a more subtle question, and to our knowledge is under-explored, particularly in the context of a single composer’s works.

\section{The Dataset}

\subsection{Basic information} 

134 different Florence Price art songs are documented to exist between the Florence B.\ Price Works Catalog \cite{price_catalog}, the Marian Anderson Collection at the Penn Libraries, and the Mullins Library of the University of Arkansas. The Mullins Library at University of Arkansas holds the largest collection of Price's music manuscripts, which can only be accessed in person upon request to the library or through photo reproduction. The Marian Anderson Collection at the Penn Libraries also contains a significant number of Price's manuscripts; for some of these, the library provides photocopies that are available online.

Of the songs, 116 are set to lyrics from a poet or by a named lyricist, including several songs which Price herself wrote the lyrics. 18 are arrangements of African-American folk songs, including spirituals. 132 of the songs are scored for a solo vocalist and a pianist. Five of the songs are incomplete.

\begin{figure}
    \centering
    \includegraphics[alt={Price's manuscript of the song \textit{I Remember}. This song demonstrates harmonic, rhythmic, and dynamic choices that characterize Price's style.},width=0.9\linewidth]{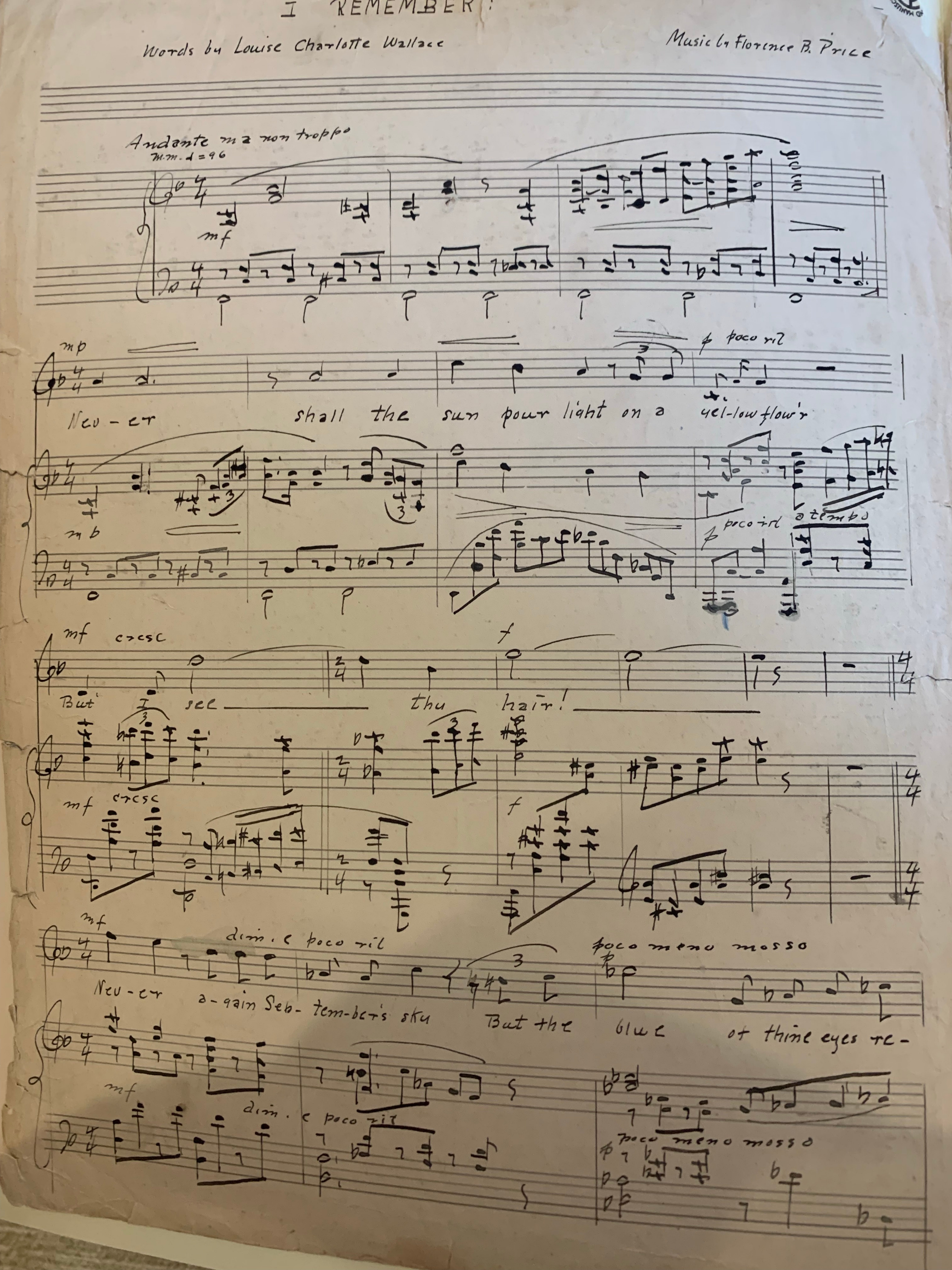}
 \caption{Price's manuscript of \textit{I Remember}, p.\ 1. \cite{uark_501224}}
    \label{FigManu}
\end{figure}

Regarding style, the songs are often highly chromatic, particularly in the piano accompaniment. Price frequently used out-of-key harmonies, diminished and augmented chords, and parallel key shifts. The whole-tone scale was a signature feature in her songs. %Pitch-class entropy \cite{jazztransformer} analysis shows that accompaniments in her songs are significantly more chromatic than those in the OpenScore Lieder corpus---see \figref{TODO}. 
Many of her piano accompaniments feature long appoggiaturas. Price used the Juba dance rhythm in many of her compositions, including many art songs. Many of her art songs include key changes and meter changes, as well as surprising rhythmic changes. For example, Price's song "I Remember" (see \figref{FigManu}) incorporated augmented fifth harmony, out-of-key chromatic notes, syncopated rhythm, activity in multiple registers, notes of a variety of durations, and a meter change. The songs are generally around 40 bars in length, with a range between 8 and 80 bars. Our project contributes to the growing analytical literature of Price's art songs \cite{unlv_2025_price_dunbar, jobson_2019_florenceprice}. 

The first author created the dataset by manually transcribing manuscripts of Price’s songs to MuseScore, and converting each MuseScore file to MusicXML, MIDI, and PDF formats. Photocopies of the manuscripts of 129 songs were obtained from the University of Arkansas and the Penn Libraries website. At first, we attempted to use Optical Music Recognition (OMR) to help digitize the manuscripts. However, OMR struggled to process the manuscripts accurately, so they were manually transcribed instead. Each transcription was reviewed twice upon completion.

\subsection{The Florence Price Art Song Dataset} 
Most of Price’s art songs entered the public domain in the United States of America on January 1, 2024, 70 years after the composer’s death in 1953. However, 17 of the 129 songs we obtained were published between 1930 and 1973, and may remain under copyright in the USA. Out of legal and ethical concerns, we exclude these 17 songs from our published dataset and the training set for our model. We release the remaining 112 songs as the Florence Price Art Song Dataset.\footnote{https://github.com/TT515/Florence\_Price\_Art\_Song\_Dataset} 

Our art song dataset includes a folder for each song. Each folder includes the MuseScore file, MusicXML, MIDI, and PDF versions of the song, as well as files for lyrics, metadata, and onsets. Metadata includes the tempo of the song, the lyricist, the first author's judgment of whether the song is happy or sad, another adjective the first author perceives to describe the song’s mood, and the song’s musical style. We provide two ``onsets'' files for each song, which denote the bar-wise section boundaries of the song as perceived by the first and second authors. We also provide a script for generating audio for every song in the dataset.

%The photocopy of the manuscript was obtained at the University of Arkansas.
%This excerpt demonstrates harmonic, rhythmic, and dynamic choices that characterize Price's style.

% \begin{figure}
%     \centering
%     \includegraphics[alt={Our transcription of the song \textit{I Remember}. The manuscript was manually transcribed into Musescore, and converted to MIDI, MusicXML, and PDF formats.},width=0.9\linewidth]{I Remember.pdf}
%     \caption{}
%     \label{fig:enter-label}
% \end{figure}

% Such boundaries refer to the start of each musical phrase, the end of the intro, and the start of the outro for each song. 

Legibility of the manuscripts presented challenges for creating the catalog. Price's intentions were not always clear, and in those instances, our transcription is based on the first author’s judgement. The author also corrected perceived errors in the manuscripts, most commonly ambiguity in accidentals. We recognize that despite efforts to present Price's compositions as accurately as possible, there may still be errors in our catalog. We welcome feedback and will actively manage the catalog to present Price's works accurately.

% \subsection{Significance} 
% This dataset fills a large gap in {\color{red}Florence Price research} by providing the first nearly complete collection of pieces in one of the musical genres in which she excelled. With access to the sheet music for these songs, more artists can record and perform them, and with a large dataset of digitized music, researchers can analyze how it compares to that of other composers, and how she approached music-lyric relationships.

% The dataset also adds to the larger corpus of available art songs, such as the OpenScore Lieder corpus \cite{OSLieder}. Incorporation of our dataset into the training data for generative models promises to add a creative touch and chromatic flair to their outputs. {\color{red} I think we can cut this paragraph if we need room.}

% \subsection{Problems, solutions}

\section{The Florence Price Piano Accompaniment Generator}
\label{SecTraining}

In this section, we describe how we created a model, hereafter referred to as the {\em FP model}, for writing piano accompaniments that reflect the style of Florence Price. Given the small size of our dataset, it would be infeasible to train such a model from scratch. Therefore, we started with the Composer's Assistant v2.1 model released in \cite{CA2} (which we refer to as the {\em starting} model) and fine-tuned it on our dataset. The starting model uses a T5 \cite{t5, huggingface} architecture and has 16 encoder layers, 16 decoder layers, 12 attention heads per layer, and a model dimension of 576, with a total parameter count of approximately 192M. 

The model accepts as input any slice of measures from a MIDI file and, for each track within the slice, a set of bar-level masks. The model then generates, using the structural information and unmasked notes in the input, notes to replace each mask. Syntactically, this is carried out with the T5 denoising objective: Simplifying the syntax of the model slightly for explanatory purposes, an example of a three-measure input to the model containing an oboe track and a piano track is [measure] [piano] [$N_{0, 0}, \ldots, N_{0, n_0}$] [oboe] [$\MASK_{0}$] [measure] [piano] [$\MASK_1$] [oboe] [$\MASK_2$] [measure] [piano] [$N_{1, 0}, \ldots, N_{1, n_1}$] [oboe] [$N_{2, 0}, \ldots, N_{2, n_2}$]. In this example, $[N_{i, 0}, \ldots, N_{i, n_i}]$ is a sequence of tokens describing the notes played by the most recent previously-declared instrument in the most recent previously-declared measure---see \cite{CA} for details.  
% {\color{red}(These tokens represent notes as collections of pitches, onset times, and durations. Each of these note properties is encoded in a one-hot fashion after quantizing to a temporal resolution of 24 ticks per beat. Tokens are listed from left to right by onset time as they occur within the measure. When multiple notes occur at the same time step, they are listed in order of pitch, from lowest to highest. See \cite{CA} for further details.)}
The oboe's part in the first measure is masked, and both parts in the second measure are masked.
The model's output would be [$\MASK_{0}$] [${M_{0, 0}}, \ldots M_{0, m_0}$] [$\MASK_1$] [${M_{1, 0}}, \ldots M_{1, m_1}$] [$\MASK_2$] [${M_{2, 0}}, \ldots M_{2, m_2}$] [eos], where $[M_{i, 0}, \ldots, M_{i, m_i}]$ is a sequence of tokens describing the notes that the model has generated to replace $\MASK_i$ in the prompt. 
Training examples are generated by taking random measure slices from the training data, masking measures within the tracks in the slice, and training the model to fill in the masked ground-truth data.  For validation and testing, we are only interested in accompaniment generation, where the model is provided with a melody and asked to write a piano accompaniment. Our accompaniment generation training objective is achieved by masking every measure in the piano track and leaving the melody track (oboe in the example) unmasked.

\begin{figure}[t]
    \centering
    \includegraphics[alt={The KL-divergence between metrics of generated accompaniments and ground-truth validation accompaniments changes with more training, but there is no clear signal as to when the model is done training.},width=0.9\linewidth]{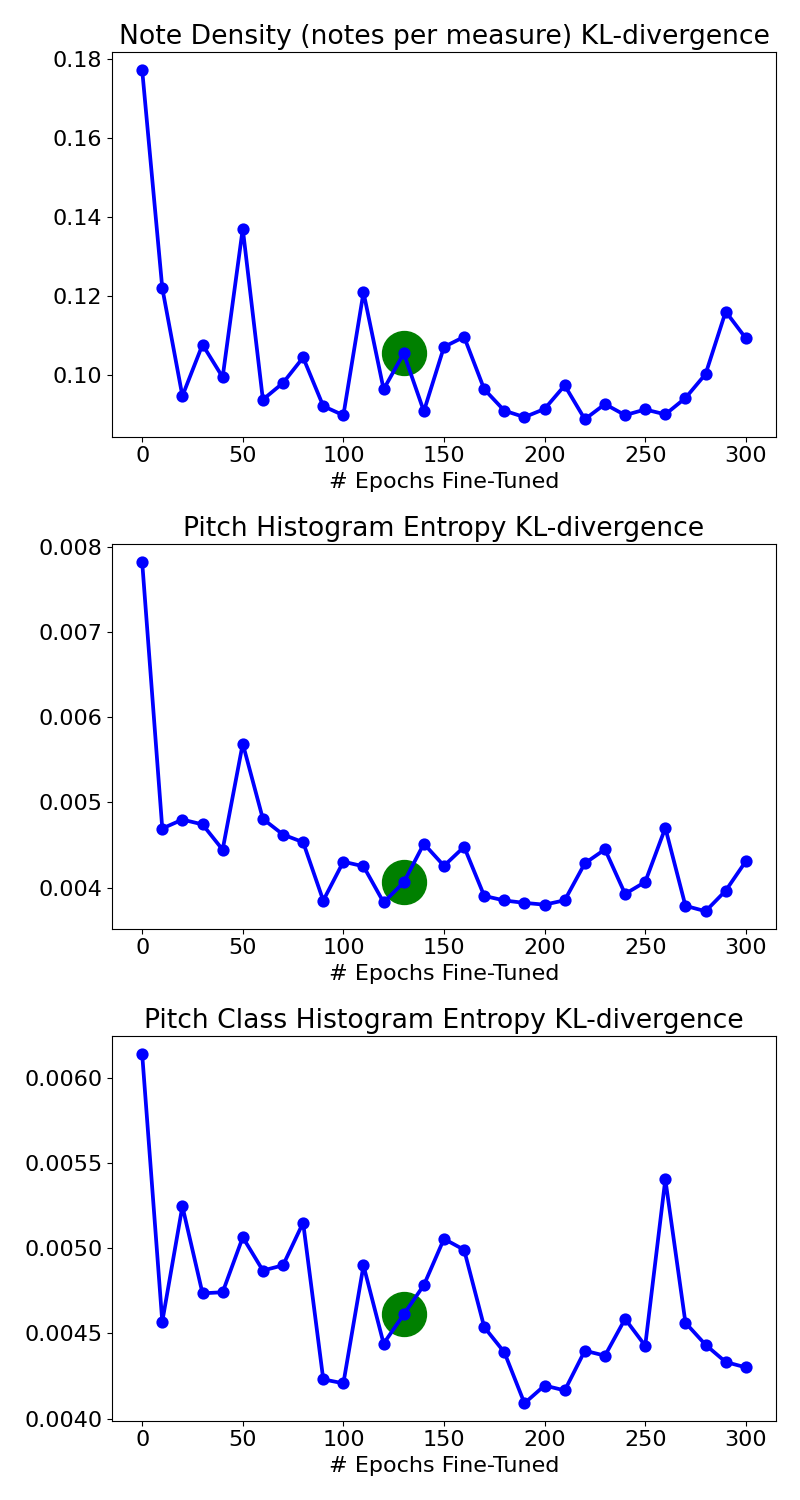}
    \caption{KL-divergence between metrics of generated accompaniments and ground truth accompaniments in the Lieder \textbf{validation} data, after fine-tuning the starting model on the Lieder Corpus. Subjectively-best model highlighted in green.}
    \label{FigFT}
\end{figure}

To create a baseline for comparison, we fine-tuned the starting model on the OpenScore Lieder Corpus \cite{OSLieder}, a collection of 1356 art songs by 19th century composers. To train this {\em baseline model}, we used 100 songs for validation and the rest for training. We used an effective batch size of 128 for training and saved a checkpoint every 10 epochs. Each epoch corresponded to about 80 optimization steps. We found the starting model unsuitable as a baseline, as it is a general model rather than an art song-specific model, and standard metrics such as note density, pitch histogram entropy (PHE), and pitch class histogram entropy (PCHE) \cite{jazztransformer} of generated accompaniments differed widely from the actual accompaniments in the validation set---see \figref{FigFT}. This figure was created by generating accompaniments to the melodies in 16-bar snippets selected at random from the validation set 
% (4 16-bar snippets per song), 
and computing the Kullback-Leibler (KL) divergence \cite{kldiv} between the metrics of generated accompaniments after $n$ epochs of fine-tuning ($n\in\{0, 10, 20, \ldots, 300\}$) and the metrics of the ground truth accompaniments. KL-divergence provides a measure of the difference between the observed and ground truth distributions. \figref{FigFT} also shows how quickly the fine-tuning process trains a model that generates accompaniments which better approximate the distributions of these metrics in the ground truth. 

\begin{figure}
    \centering
    \includegraphics[alt={The pitch histogram entropy of generated accompaniments to melodies in the training set nearly matches that of the ground truth after 60 epochs of training, indicating memorization of training data.},width=1\linewidth]{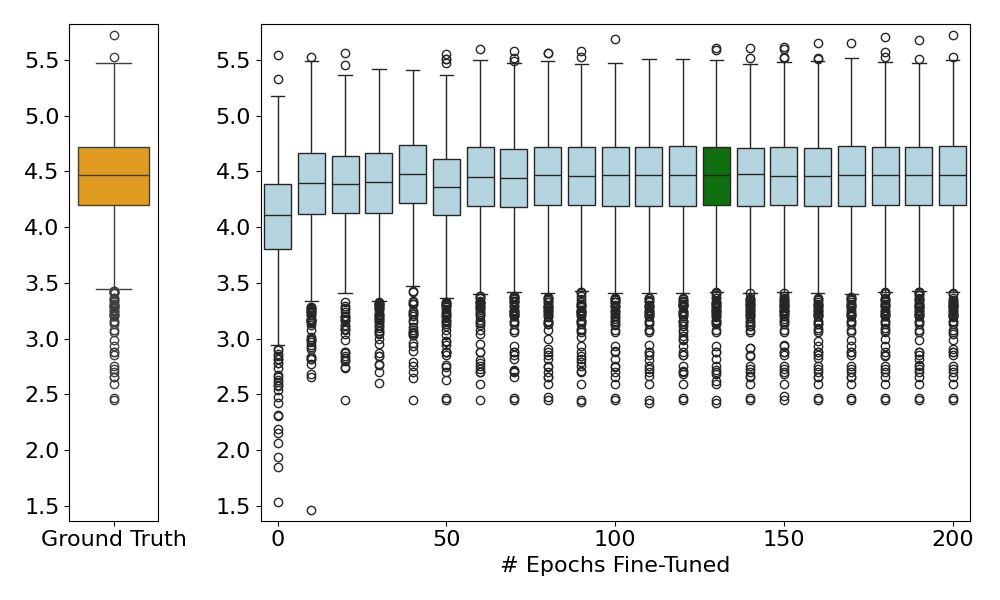}
    \caption{Pitch histogram entropy of generated accompaniments to melodies in the Lieder \textbf{training} data, after fine-tuning the starting model on the Lieder Corpus. Subjectively-best model highlighted in green.}
    \label{FigValEqualsTrain}
\end{figure}

Despite the popularity of fine-tuning generative creative models on small datasets for style purposes (e.g., in art and music), there does not appear to be much guidance in the literature about {\em exactly} when to stop the fine-tuning process. Previous works that fine-tuned symbolic music models include \cite{ProgGP, ShredGP, tradformer, jigs}. The most common guidance given is to save a training checkpoint every so often, examine some model outputs from the checkpoint, and decide whether to keep training. Eventually, training is stopped when the outputs noticeably degrade as the model overfits. Ultimately, this was our methodology for training both the baseline model and the FP model. 
In one notable exception, the authors of \cite{ProgGP} found success in comparing the PHE of generated samples to the PHE of the training data to decide when to stop fine-tuning their model---they observed divergence between these quantities as their model overfit. We observed the opposite signal while training our baseline model, both when comparing to the validation data (center plot of \figref{FigFT}) and to the training data itself (\figref{FigValEqualsTrain}), so this did not provide a clear signal as to when to stop training. Rather, the convergence in \figref{FigValEqualsTrain} is consistent with the model {\em memorizing} its training data. Some memorization is acceptable, as long as the quality and diversity of generated accompaniments to melodies outside the training set do not suffer. 

As noted in \cite{DiscreteDiffusionSymbolic}, metrics commonly used for the evaluation of generated music---both objective and subjective---tend to focus on quality rather than diversity of outputs. We want our trained models to be able to create a wide variety of high-quality accompaniments for any given input melody. For inference, we use nucleus sampling \cite{nucleussampling} with $p=0.95$ and a temperature of 1.0, and we found that increasing {\em rhythmic temperature} to 1.5 during inference helps maintain a diversity of outputs and allows the models to train for more epochs, leading to higher-quality outputs. (The model generates accompaniments one token at a time, where a token controls the pitch, position, or duration of a generated note. By rhythmic temperature, we refer to the temperature parameter when the model is choosing the next position or duration token, but not when it is choosing the next pitch token.)

To train the FP model, we fine-tuned the starting model on our Florence Price Art Song Dataset and saved a checkpoint every 10 epochs. Due to the small size of our dataset, each epoch corresponded to about 2.7 training steps. By listening to accompaniments generated to melodies outside the training sets, we determined that training the baseline model for 130 epochs and training the FP model for 100 epochs was optimal. 

While training, we examined validation loss, note density, PHE, PCHE, and F1, both on the validation data (in the case of the baseline model) and on the training data, and found no clear signal in any of them pointing us towards the model checkpoints we perceived as optimal. We therefore call for more research into stopping criteria for style-based fine-tuning on small datasets. The only clear signals we found that pointed us in the right direction were: (1) a noticeable drop in the diversity of generated accompaniments to given melodies after too much training, and (2) {\em catastrophic forgetting} \cite{Cata}: The starting model comes equipped with control tokens that can be supplied in prompts to affect the musical properties (such as note density) of the generated outputs, and eventually these control tokens become increasingly ineffective (i.e., ignored by the model) with additional training. Both issues began to occur at 150 epochs with our FP model and at 170 epochs with the baseline model, and in both cases were exacerbated by more training, giving us an upper bound for our listening-based search.

% We added a ``last bar'' token to the vocabulary of both models (which was not present in the starting model) to help them learn how to write endings to pieces. During training, this token is inserted immediately after the [measure] token for the measure containing the last note onset in the song. During inference, this token is inserted into the last measure of prompts to tell the model to write an ending. Without this token, due to the imbalance in training data (the end of a phrase is usually not the end of the song), the models would frequently attempt to write a transition to an anticipated next part of the song instead of an ending. We demonstrate in \secref{SecEval} that this token---whose meaning was learned by the models entirely during the fine-tuning process---is generally effective in signaling to the model to write an ending rather than a transition.

We release the FP model and code for interacting with it as the Florence Price Piano Accompaniment Generator. We provide an interface where a user can upload any melody in MIDI format and download piano accompaniments generated by the model.

% For evaluation (see \secref{SecEval}), since our dataset is small, we trained 10 additional Florence Price models by leaving one song out of our training dataset---for these models, based on our findings above, we simply trained for 100 epochs. All of our training experiments were performed on a single NVIDIA® GeForce RTX™ 4090 GPU. Total training time for all experiments was 110 hours, of which the final fine-tuning run to create our model took 75 minutes.

\section{Evaluation}
\label{SecEval}
Due to the novelty of the dataset presented in this paper, it would be difficult to find external evaluators who are familiar with the style of the music in the dataset. Therefore, the first and third authors, who have listened to all of the music in the dataset, participated as evaluators in a blind listening experiment to assess the extent to which the FP model (developed in \secref{SecTraining}) is perceived as writing accompaniments that are reflective of Florence Price's style.
%The experiment assesses the extent to which the FP model developed in \secref{SecTraining} is perceived as writing accompaniments that are reflective of Florence Price's style. 
% Due to the novelty of the dataset and the rarity of people with expertise in Price's art songs, it was hard to find suitable people to add to the evaluation process aside from the authors.
The second author, who performed model checkpoint selection in \secref{SecTraining}, used the FP model and the baseline model to generate accompaniments to specific melodies for the evaluators to listen to. No control tokens from the starting model (e.g., note density, pitch range) were used, in order to give the models the most flexibility in their generation. 

% However, we included the ``last bar'' control token introduced in \secref{SecTraining} in the last measure of every prompt.

We selected 10 melodies from Florence Price's art songs and 10 popular melodies, and for both models, generated 7 accompaniments to each melody. This created 140 sample accompaniments from each model. To generate the accompaniments to Price's melodies, we did not use the fine-tuned model from \secref{SecTraining}. Rather, we employed a leave-one-out training approach, training the starting model for 100 epochs on our dataset without the song containing the melody, and using the resulting model to generate the accompaniments for that melody. 
% Each such model took 25 minutes to train. 

The evaluators then blindly and separately listened to pairs of accompaniments to each melody (one accompaniment generated by the FP model, the other by the baseline model) and selected the accompaniment they felt better reflected the style of Florence Price. The evaluators listened to the same 280 samples, but paired in different ways. These samples were neither edited nor cherry picked. These examples and many more (1400 in total) are available for the reader to listen to.\footnote{https://github.com/m-malandro/Florence-Price-listening-examples}

We give the results of this experiment in \tabref{tab:subjective}. The evaluators considered accompaniments generated by the Florence Price model to be reflective of Price's style more often than accompaniments generated by the baseline model for both types of melodies.
% , indicating successful capturing of her accompaniment style on her own melodies as well as style transfer to melodies of a different nature from those in our training data. 
% However, {\color{red} say something about how subtle you found her style to be expressed in some of the popular melodies? idk - this is up to you} {\color{blue} would like to hear Dr Shadle's take on the generations}

\begin{table}[t]
    \centering
    \begin{tabular}{|c|c|c|c|}
    \hline
      Melody type & $n$ & Price model wins & $p$-value\\
      \hline
      \hline
       Price  & 137 & 82 & $0.026$\\
       \hline
       Popular  & 137 & 99 & $<0.001$ \\
       \hline
    \end{tabular}
    \caption{Results from our blind listening experiment. The listeners considered accompaniments generated by the Florence Price model to be reflective of Price's style more often than accompaniments generated by the baseline model. Six out of 280 responses were missing. $p$-values were obtained from binomial tests.}
    \label{tab:subjective}
\end{table}

\begin{table}[t]
    \centering
    \begin{tabular}{c|c|c|}
     % & Florence & Lieder \\
     % & Price & (baseline) \\
     & Our & Baseline \\
     & Model & Model \\
     \hline
    Hard errors &  67 & 47\\
    Soft errors & 68 & 57 \\
    \hline
    \% accompaniments & \multirow{2}{*}{45.71\%} & \multirow{2}{*}{57.14\%}\\
    with no errors & &\\
    \hline
    \end{tabular}
    \caption{Perceived errors in generated accompaniments (140 from each model). Each accompaniment was 20-60 seconds, averaging 30 seconds, and was independently examined by two listeners. Both listeners agreed that 45\% of the clips generated by the FP model had no errors.}
    \label{tab:errors}
\end{table}

\begin{figure*}
    \centering
    \includegraphics[alt={The first 9 measures of generated accompaniments to Rhapsody},trim={0 30 0 0},width=\linewidth]{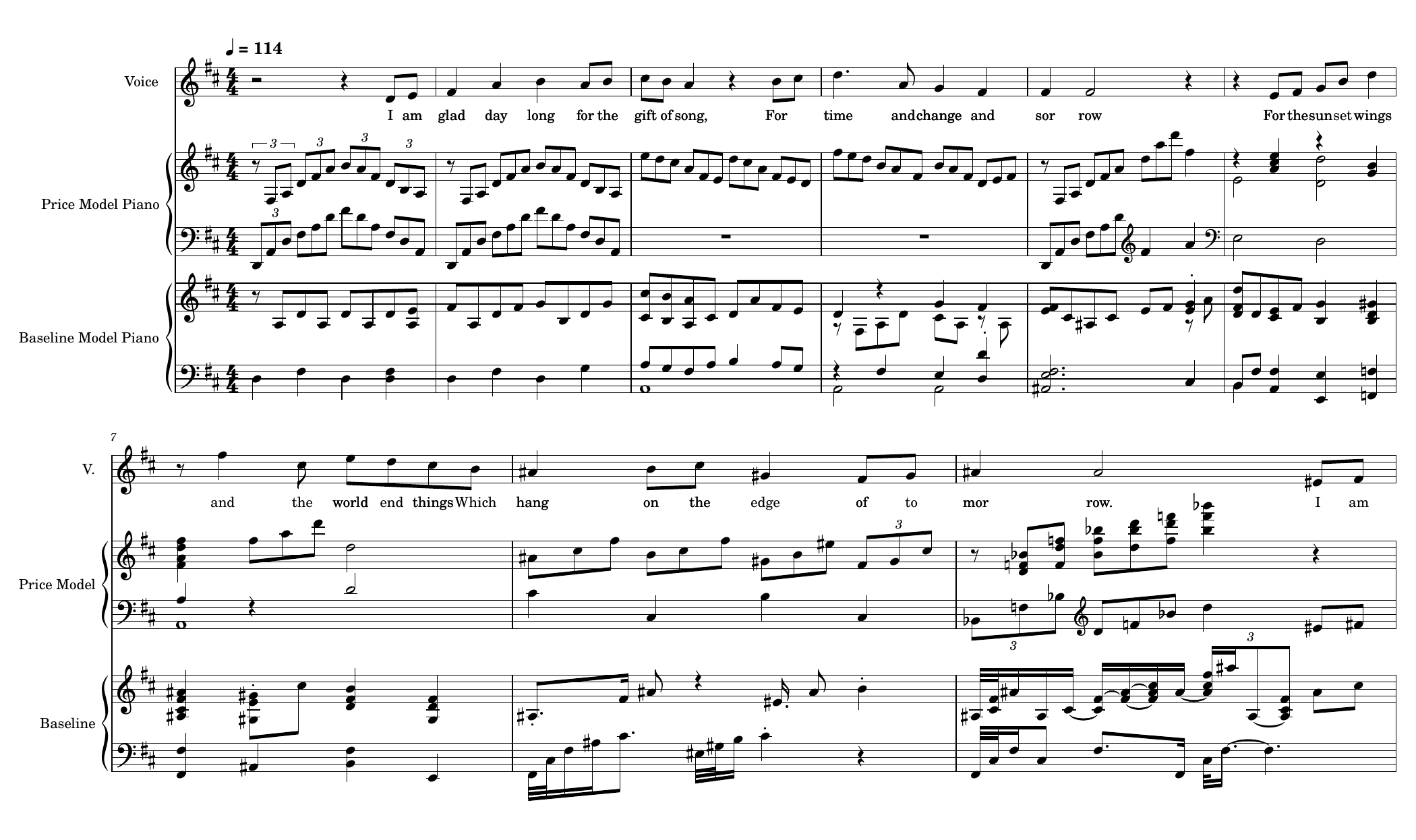}
    \caption{First 9 measures of generated accompaniments to {\em Rhapsody}.}
    \label{fig:compared}
\end{figure*}

% \begin{figure}[h]
% \includegraphics[alt={The melody of \textit{Rhapsody} with accompaniment generated by the baseline model.},width=0.9\linewidth]{Rhapsody Lieder1.mid.pdf}
% \caption{The melody of \textit{Rhapsody} with accompaniment generated by the baseline model.}
% \label{FigManu}
% \end{figure}

% \begin{figure}
% \includegraphics[alt={The melody of \textit{Rhapsody} with accompaniment generated by the Florence Price model.},width=0.9\linewidth]{Rhapsody FP0.pdf}
% \caption{The melody of \textit{Rhapsody} with accompaniment generated by the Florence Price model.}
% \label{FigManu2}
% \end{figure}

% It was clear during training that our ``last bar'' token was effective. To evaluate its effectiveness here, the second author took 50 of the generated accompaniments at random (25 from the baseline model and 25 from the Florence Price model) and 50 accompaniments generated by the starting model (which does not have the ``last bar'' token in its vocabulary), and rated whether the accompaniments had a satisfying ending. 

As a secondary objective, to help assess the quality of the FP model and its potential for use in co-creative composition, the evaluators took notes of how many ``hard'' and ``soft'' errors they heard in each generated accompaniment. These counts are presented in \tabref{tab:errors}. For purposes of this paper, a hard error is a significant occurrence of harmonic dissonance or melodic incoherence, while a soft error is an occurrence of coherent but counter-intuitive melodic or harmonic sequences, such as an unprompted use of inversions or accidentals. Sometimes the output contains no errors as defined, or only soft errors that could be fixed by a skilled composer.

% {\color{}Generative symbolic music models make mistakes. 
% Based on the harmony and melody, we define two types of errors, hard errors and soft errors, for this paper.  When the output contains hard errors, it may need to be completely regenerated. 
% \color{red} Composers who use generative models in a co-creative fashion have multiple roles, including prompt engineer, curator, and editor.
% We do not test whether the FP model is as good as a human composer---we know it is not, and we are not aware of any study claiming that any generative symbolic model is capable of consistently generating results that are of human-level quality. 
% However, some studies have found that cherry picking \cite{betweenAIandMe} and editing \cite{CA2} outputs can come close to rivaling human-composed music. }

The baseline model generates error-free accompaniments 
% significantly $(p=0.028)$ 
more often than the FP model, which we trace back to the small size of our fine-tuning set. 
However, for co-creative applications, the goal is not necessarily perfection, but rather providing stylistically appropriate and usable material. While an error-free accompaniment is not necessarily a {\em good} accompaniment, the vast majority of our generated accompaniments, including those with errors, feature musically interesting rhythmic textures, harmonic voicing, movement between registers, and an appropriate amount of embellishments. These are all standout features of the art song genre. Our perceived error-free accompaniment generation rate of 45\% therefore suggests that the FP model can meaningfully support composers seeking inspiration in Price’s style.

Finally, as a case study, we compare two accompaniments generated to the melody of {\em Rhapsody}, composed by Florence Price. One was generated by the baseline model, and the other was generated by our model after fine-tuning on the catalog of Price's songs excluding {\em Rhapsody}. The first nine (out of 22) bars of the two accompaniments are given in \figref{fig:compared}. The two 22-bar samples are available for the reader to listen to.\footnote{https://github.com/m-malandro/Florence-Price-listening-examples/releases/download/v1.0.0/Rhapsody.zip} 

We analyze the rhythm, range, and harmony of the two accompaniments. The baseline model's rhythm is based on eighth notes, and deviates to a more active rhythm featuring 32nd notes at bar eight. The FP model's rhythm is based on triplets and deviates to a quarter-note rhythm in bar six, before returning to triplets and reaching a climax at bar nine with ascending triplet chords. Florence Price often uses fast ascending chords in the accompaniment to reach climaxes in her songs, so this is an example of the FP model incorporating Price's musical language into its generations. The FP model sample is active in a wider range, similar to many of Price's art songs accompaniments. Both accompaniments feature harmonies coherent with the context of the melody and the accompaniment's own progression, although they differ in bar nine, where the baseline model's harmony is F\# major and the FP model's harmony is B\ensuremath{\flat} major (treating the A\# in the melody as a B\ensuremath{\flat}). B\ensuremath{\flat} major is more surprising and expressive in the context of the home key D major, and although Price's original composition had F\# major harmony in bar nine, the B\ensuremath{\flat} major harmony is encouraging because it reflects Price's tendency to use surprising and expressive harmony in her accompaniments. By comparing the harmony, rhythm, and range of the sample accompaniments, we showcase the FP model's capacity to generate accompaniments that align with Price's musical vocabulary and expressive tendencies.

\section{Conclusion}

We have created and released a digital catalog of Florence Price's works for solo voice and piano in MuseScore, MusicXML, MIDI, and PDF format. Using the catalog, we have fine-tuned and released a symbolic music model to generate piano accompaniments reflective of Price's compositional style. We have conducted a listening experiment and concluded that the accompaniments generated by the fine-tuned model successfully captured elements of Price's style. We hope that our digital catalog can serve as a pillar for artists, listeners, and researchers interested in learning more about Price's songs, and we hope that our model can serve as a tool both for composers and those interested in analyzing Price's music. 

% , as well as an example of fine-tuning on a single composer's work.

% \section{Acknowledgments}
% We would like to thank The David W. Mullins Library and The Penn Library for allowing access to Price's manuscripts. 

\section{Ethics Statement}

All of the music we have released in the Florence Price Art Song Dataset and used to train the Florence Price Piano Accompaniment Generator entered the public domain in the United States of America on Jan.\ 1, 2024. We obtained access to Price's manuscripts with permission from the University of Arkansas David W. Mullins Library and the University of Pennsylvania Kislak Center for Special Collections, Rare Books, and Manuscripts.

% In this work, we have explored fine-tuning a model to enable the model to write music in the style of a particular composer. While our findings can benefit composers who wish to fine-tune models to write in their own styles for creative or co-creative purposes, there is also the risk that our work may inspire composers to fine-tune models to impersonate the styles of others.

% For BibTeX users:
% \bibliography{ISMIRtemplate}
\bibliography{main.bbl}

\newpage
\clearpage
\newpage

\end{document}